\documentclass[journal]{IEEEtran}
\IEEEoverridecommandlockouts
\usepackage{cite}
\usepackage{amsmath,amssymb,amsfonts}
\usepackage[linesnumbered,ruled]{algorithm2e}
\usepackage{graphicx}
\usepackage{textcomp}
\usepackage{xcolor}
\usepackage{epsfig}
\usepackage{multirow}
\usepackage{subfigure}

\usepackage{svg}

\def\BibTeX{{\rm B\kern-.05em{\sc i\kern-.025em b}\kern-.08em
    T\kern-.1667em\lower.7ex\hbox{E}\kern-.125emX}}
\usepackage{cite}
\begin{document}
\bstctlcite{IEEEexample:BSTcontrol}
\title{A 1.6-mW Sparse Deep Learning Accelerator for Speech Separation}

\author{Chih-Chyau Yang,~\IEEEmembership{Associate Member,~IEEE},~Tian-Sheuan~Chang,~\IEEEmembership{Senior Member,~IEEE} %
\thanks{Manuscript received September 21, 2022; revised December 10, 2022, and accepted January 6, 2023.}
\thanks{This work was supported by the Ministry of Science and Technology, Taiwan, under Grant 111-2622-8-A49-018-SB, 110-2221-E-A49-148-MY3, and 110-2218-E-A49-015-MBK. 
Chih-Chyau Yang is with the Institute of Electronics, National Yang Ming Chiao Tung University, Taiwan, and the Taiwan Semiconductor Research Institute, Taiwan. (e-mail: ccyang@narlabs.org.tw)
Tian-Sheuan Chang is with the Institute of Electronics, National Yang Ming Chiao Tung University, Taiwan. (e-mail: tschang@nycu.edu.tw)}}
\maketitle

\begin{abstract}
Low power deep learning accelerators on the speech processing enable real-time applications on edge devices. However, most of the existing accelerators suffer from high power consumption and focus on image applications only. This paper presents a low power accelerator for speech separation through algorithm and hardware optimizations. At the algorithm level, the model is compressed with structured sensitivity as well as unstructured pruning, and further quantized to the shifted 8-bit floating-point format instead of the 32-bit floating-point format. The computations with the zero kernel and zero activation values are skipped by decomposition of the dilated and transposed convolutions. At the hardware level, the compressed model is then supported by an architecture with eight independent multipliers and accumulators (MACs) with a simple zero-skipping hardware to take advantage of the activation sparsity and low power processing. The proposed approach reduces the model size by 95.44\% and computation complexity by 93.88\%. The final implementation with the TSMC 40 $nm$ process can achieve real-time speech separation and consumes 1.6 mW power when operated at 150 MHz. The normalized energy efficiency and area efficiency are 2.344 TOPS/W and 14.42 GOPS/mm$^2$, respectively.

\end{abstract}


\IEEEpeerreviewmaketitle

\section{Introduction}
Separation of speech, music and environmental sounds is an important task for many speech applications and automatic machine hearing, such as the automatic speech recognition and music applications in edge devices\cite{mirbeygi2022speech}. Its quality has been significantly improved with the introduction of deep learning. However, the deep learning models suffer from high computational complexity and inhibit its low power real-time execution at the edge devices, which cannot be met by power hungry CPUs and GPUs. To meet low power and real-time requirement, a promising approach is the hardware accelerators\cite{sze2017efficient}. Although many deep learning accelerators \cite{eyeriss, eyerissv2,vwa,HsiaoDLA} have been proposed, most of them are designed for image applications, and one is for speech enhancement\cite{NTUADSP} used in hearing aids by frequency domain processing. These designs are for 2-D image or image like processing and do not take full advantages of 1-D speech processing and thus suffer from the low hardware utilization problem.

This paper presents a low power design for speech separation with algorithm level complexity reduction and hardware optimization for sparse processing. The algorithm level exploits speech-related features for complexity reduction approaches like structured sensitive pruning and unstructured pruning, as well as shifted floating-point quantization. Furthermore, with the decomposition for the dilated and transposed convolutions, computations with the zero kernel and zero activation values are skipped. These reduce the complexity by 89.99\%. At the hardware level, an eight independent 8-bit floating-point MAC architecture is proposed to support the compressed model. The computations with zero activations in 1-D convolution and 1×1 pointwise convolution are skipped by a simple zero-skipping hardware with broadcasting mechanism while fully utilizing the hardware. This further reduces the complexity by 38.82\%. These speech related optimizations achieve 
1.6 mW low power consumption for real-time hardware implementation.

 The rest of the paper is organized as follows. Section II introduces the related works. Section III illustrates the proposed model compression technique and decomposition scheme. Section IV presents the proposed deep learning accelerator (DLA) architecture, the zero skipping hardware and the data flows for the separation model in hardware. The implementation results and design comparisons are shown in Section V. Finally, Section VI concludes the paper.
 
\section{Related Works}
\subsection{Speech separation}
A time-domain fully-convolutional audio separation network \cite{convtasnet,tasnetexp} has been proved to obtain better performance than in short-time Fourier transform (STFT) domain, because it resolves disadvantages of the high latency of calculating the STFT, and the decoupling of phase and magnitude in STFT. A separation network with multi-resolution encoder and decoder is further proposed to achieve better performance than a single resolution time-domain separation network in singing voice separation application\cite{chi_separation, chi_melody}.

\subsection{Model compression}
Many model compression techniques\cite{hansong2020, speechcompression, radosavovic2020designing, logBFP, cambier2020shifted, SmaQ, hybridfp8, our_dt, 2019rethinkingpruning} have been presented for deep learning models to reduce memory size and computational complexity. The model compression techniques can be classified into two approaches: pruning and quantization. 

The pruning technique often includes structured pruning and unstructured pruning schemes to remove unimportant weights \cite{hansong2020, speechcompression}. The structured pruning technique performs the filter, channel and layer pruning, while the unstructured pruning technique performs the element-wise pruning. 
In the case of structured pruning, the network can either be pruned using a predefined target network and directly trained the small target model from random initialization, or the pruned network model can also be auto-discovered and trained from scratch\cite{2019rethinkingpruning}. In the case of the unstructured pruning, the unstructured pruned model can be obtained from the iterative procedures of training, element-wise pruning, and fine-tuning.

The second compression technique is the quantization of the model. The quantization technique reduces the number of bit widths for weights and activations and thus achieves the low requirements of memory bandwidth and SRAM storage \cite{logBFP, cambier2020shifted, SmaQ, hybridfp8}. With the appropriate selection of quantization data format and data size, the retraining network model can maintain comparative performance. The numerical representations of a quantized value can be divided into two classes. The fixed-point representation possesses the less dynamic range, but an arithmetic unit using the fixed-point representation is more area-efficient, while the floating-point representation owns the larger dynamic range, but the floating-point arithmetic unit may consume more area overhead. When using the floating-point format, only fewer bits are needed to represent a specified value, and thus gain the benefit from the reduction of memory bandwidth and memory storage. In addition to the model compression technique, the decomposition scheme can be used to reduce the computational complexity of a 2-D network model \cite{our_dt}. The operations such as the depthwise dilated convolution and transposed convolution may introduce many zeros computations. The decomposition technique enables one to skip the zero computations and reduce the computation complexity.

\subsection{Deep learning accelerators}
The current existing deep learning accelerators\cite{eyeriss, eyerissv2,vwa,HsiaoDLA, NTUADSP, 2021issccasr} are usually designed to meet high-performance needs with large number of processing elements and different data flows. However, these state-of-the-art hardware accelerators have low hardware utilization while executing the time-domain separation model including the four main key operations of the 1-D convolution, the 1-D depthwise dilated convolution, 1-D 1×1 convolution and 1-D transposed convolution. In addition, these hardware accelerators are not suitable for sparse execution either.

\section{The Model Compression and Decomposition}

\subsection{Baseline model and its complexity analysis}

\begin{figure}[t]
	\centering{\includegraphics[width=0.45\textwidth]{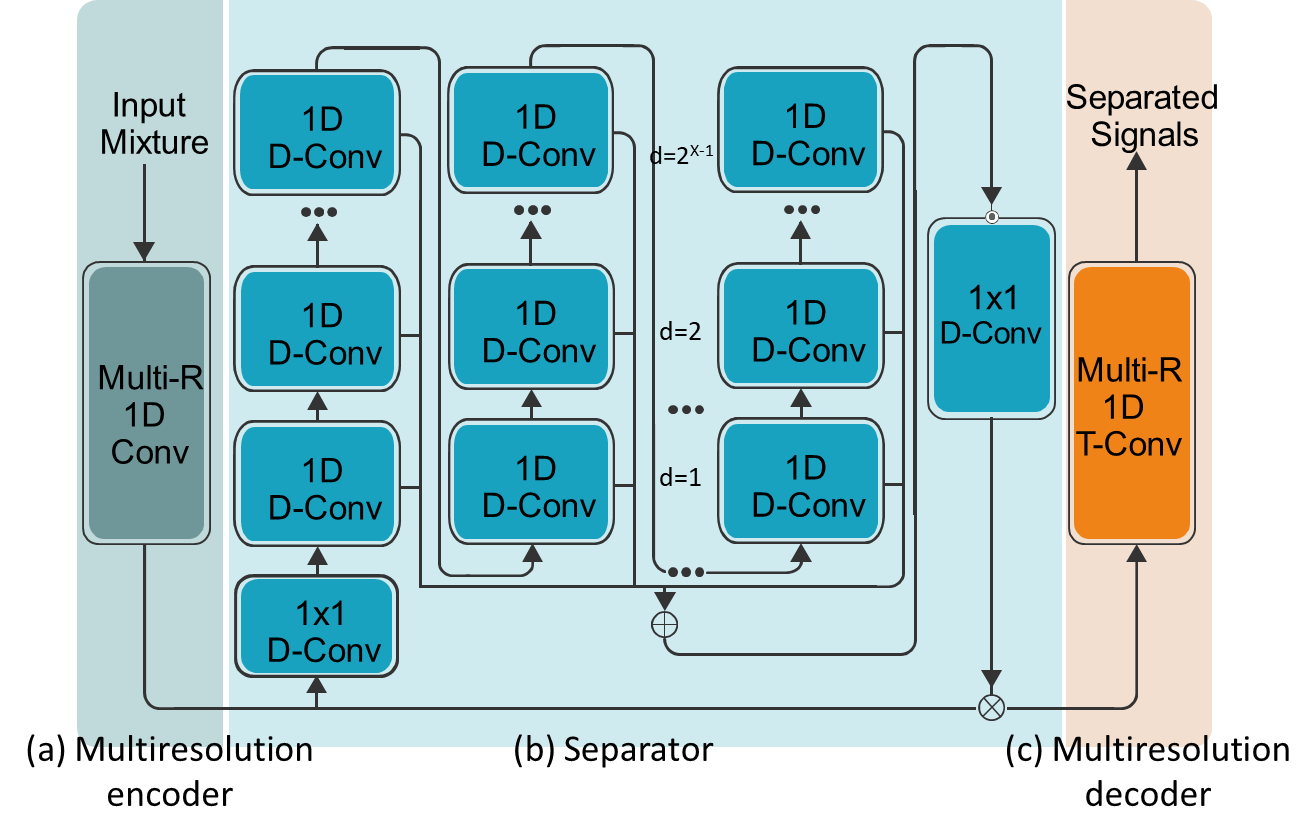}}
	\caption{The Multiresolution separation network including an encoder, a separator, and a decoder}
	\label{fig:tasnet}
\end{figure}

Fig.~\ref{fig:tasnet} shows a state of the art speech separation neural network model \cite{chi_separation, chi_melody} used in this paper, which consists of a multiresolution encoder, a separator, and a multiresolution decoder. The encoder and decoder, inspired by human hearing perception, use several short- and long-length 1-D filters to produce multiresolution spectrogram-like graphs without suffering from the trade-off between spectral and temporal resolutions. The separator uses a series of 1-D depthwise dilated convolutions to enlarge the receptive fields without significant complexity increase, and generates signal masks for signal separation. The masks are multiplied with the encoder output to select the desired signals, and its result is reconstructed to two separated signals by the decoder with the transposed convolutions.

\begin{figure}[t]
	\centering{\includegraphics[width=0.45\textwidth]{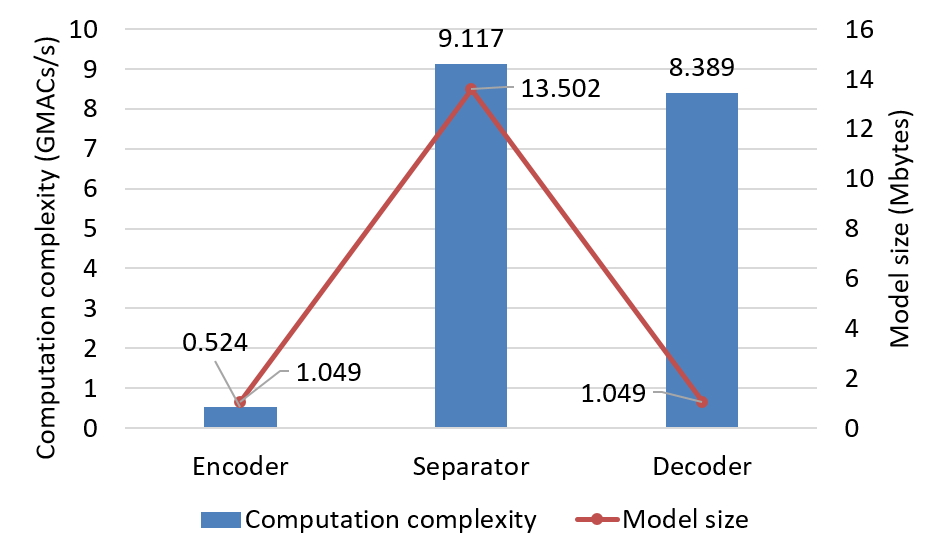}}
	\caption{The computation complexity and weight size of the baseline model}
	\label{fig:baseline}
\end{figure}

Fig.~\ref{fig:baseline} shows the complexity analysis of the baseline model for 1-sec speech data. The total size of the model is 15.6 MB, where the encoder and decoder occupy 13.45\%, and the separator occupies 86.55\%, respectively. The computational complexity is 17.99 GMACs/s (Giga MACs per second) for 16 KHz input signals, where the encoder and decoder occupy 49.52\%, and the separator occupies 50.48\%, respectively. External memory access to execute this model is 342.27 MB/s. With such a high complexity, it is difficult to deploy this model on edge devices.

\subsection{Overview of the proposed approach}

To enable low power and real-time execution of speech separation at the edge devices, model compression is first applied to reduce model size and complexity. This paper adopts mixed-level model pruning that uses structured sensitivity pruning and unstructured pruning to reduce the size of the model and the complexity of the computation. The pruned model is further quantized from the 32-bit floating-point format to the shifted 8-bit floating-point format with little performance degradation. In addition, to avoid computing with zero values in the dilated and transposed convolutions, we apply a 1-D decomposition scheme to these convolutions to remove the computations with zero weights and zero activations, respectively.

\subsection{Pruning}
\label{subsection:Pruning}

\begin{figure}[t]
	\centering{\includegraphics[width=0.48\textwidth]{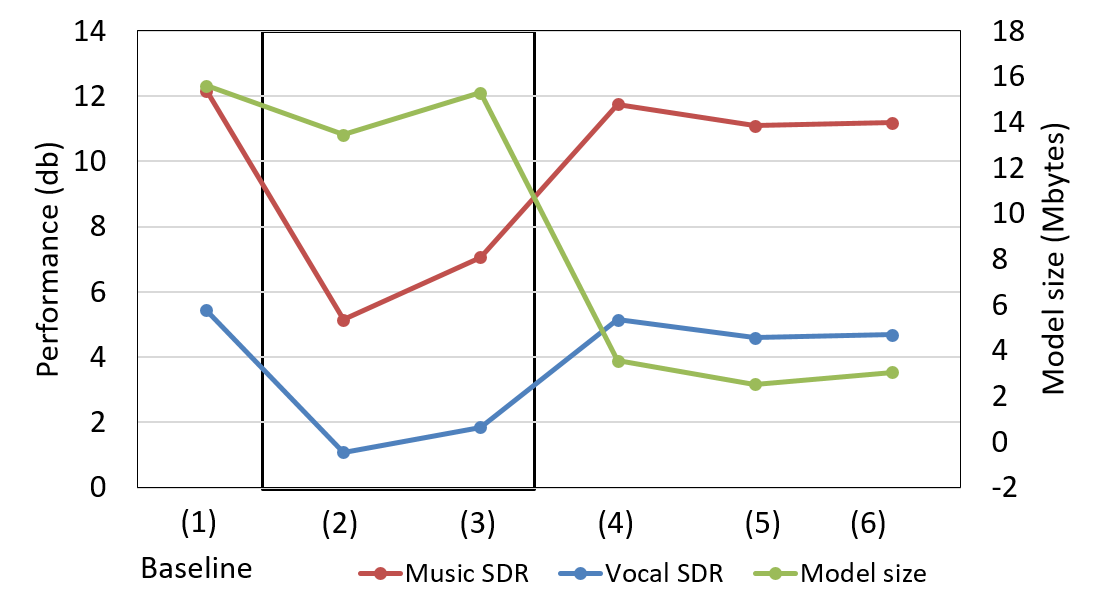}}
	\caption{The initial results of the structure pruning by shrinking the layer and channel numbers. The notations (1)-(6) represent the parameters of (2, 512, 8, 3, 512), (1, 512, 8, 3, 512)*, (1, 512, 8, 3, 512)**, (2, 256, 8, 2, 128), (2, 256, 8, 1, 128), and (2, 256, 4, 2, 128), respectively. For the network parameters of $\{L, N, X, R, H\}$, $L$ indicates the layer number in encoder and decoder; $N$ indicates the channel number in encoder, decoder and depthwise dilated groups of separator; $X$ means the length for a depthwise dilated group; $R$ represents the repeat number of depthwise dilated groups; $H$ indicates the channel number in 1×1 pointwise convolution block. Note that * indicates the long 1-D filter layer is pruned, and ** indicates the short 1-D filter is pruned. }
	\label{fig:test-accuracy-temp}
\end{figure}

\subsubsection{Structured sensitive pruning}
Fig.~\ref{fig:test-accuracy-temp} shows the initial results of the structured pruning by shrinking the layer and channel numbers. This figure shows that the layer pruning in the encoder and decoder networks is more sensitive than that in the separator network due to its regression computation nature. In contrast, the number of layers and channels in the separator network can tolerate a higher pruning ratio due to its classification nature. Thus, we apply the channel pruning in the encoder and decoder and both layer and channel pruning in the separator network, which is denoted as structured sensitive pruning.

The proposed structured sensitivity pruning flow is described below. To simplify the pruning process, we adopt the predefined structured pruning technique \cite{2019rethinkingpruning} that directly shrinks the model structure to the target size and trains from scratch. In this paper, based on the network sensitivity, we gradually shrink the channel width of the encoder and decoder, and remove the layer and shrink channel width of the separator until the target size and accuracy constraints are met. The pruned network is also the input of the unstructured pruning algorithm.

\subsubsection{Unstructured pruning}

After the structured sensitivity pruning, we further reduce the weight size by the unstructured pruning. The unstructured pruning flow follows the approach in \cite{speechcompression} and is described below. The algorithm first sets the target test accuracy and model size as constraints, and the sparsity threshold for pruning. In the first iteration, the model weights are pruned if their absolute values are smaller than the threshold value, and then the pruned network is retrained to meet the constraint.  The  threshold value is increased, and the iteration is repeated until the target constraints are met. Unlike structured pruning that will have a regular network structure, unstructured pruning will cause an irregularity in weight distribution. However, the sparsity ratio is low since most unimportant weights have been pruned by structured sensitive pruning.

\subsection{Shifted 8-bit floating-point quantization}
\begin{figure}[t]
	\centering{\includegraphics[width=0.35\textwidth]{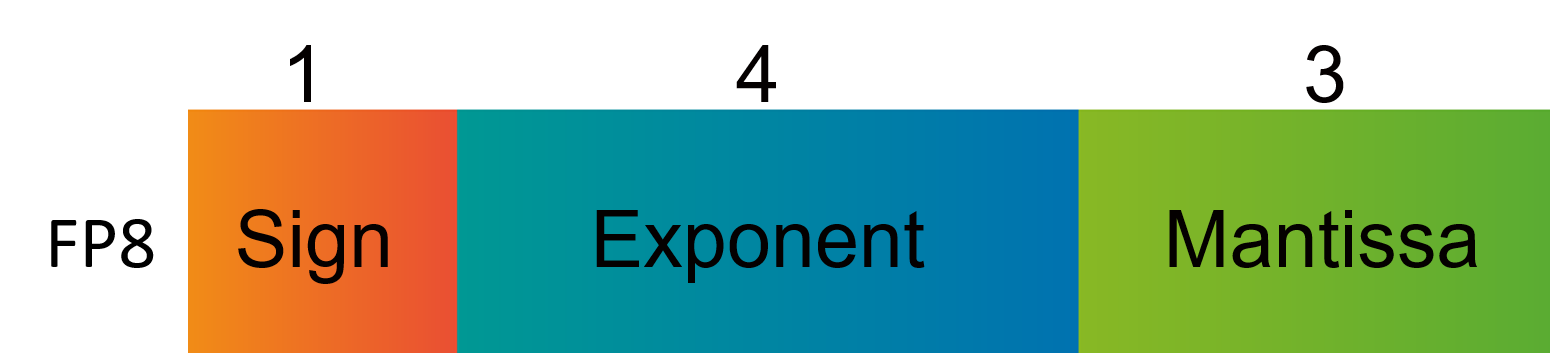}}
	\caption{The proposed 8-bit floating point format with new bias}
	\label{fig:floating-format}
\end{figure}

The model quantization technique reduces the number of bit width for the weights and activations, and thus lowers the requirements of memory bandwidth and SRAM storage. From the application viewpoint, the dynamic range requirement (16-bit data or more) in the speech application is much wider than that in image applications (8-bit data). Thus, how to get a good trade-off between area cost and dynamic range is important for the hardware design. To meet required dynamic range with lower hardware cost and power, we adopt the shifted 8-bit floating-point for weight and activation quantization. Designs with a low bit-width floating-point have a small area overhead, but with comparative performance as its 32-bit counterpart. The floating-point number $F$ can be represented as: 
\begin{equation}
F {=}{(-1)}^S\times1.M\times2^{(E-B)}
 \label{eq:float}
\end{equation}
\noindent where $S$ indicates the sign bit, $E$ means the exponent value, $M$ represents the fraction value, and $B$ indicates the bias value. The bias $B$ for the floating-point number is equal to $2^{(A-1)}-1$, where $A$ indicates the number of exponent bits $E$. The format of the 8-bit floating-point number with exponent number 4 is shown in Fig.~\ref{fig:floating-format}, with bias equal to 7.

\begin{figure}[t]
	\centering{\includegraphics[width=0.40\textwidth]{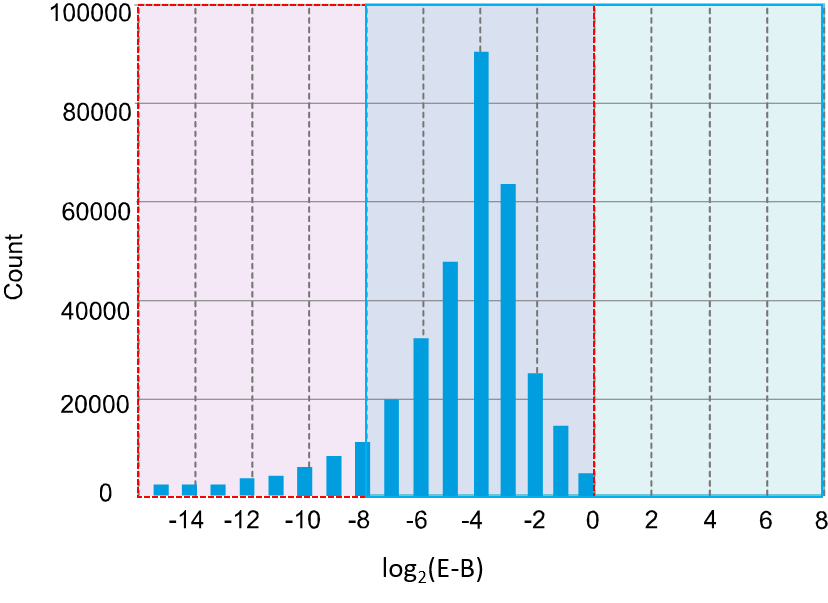}}
	\caption{The dynamic range is shifted to cover most distributions of weights and activations for the target model}
	\label{fig:weight-dynamic-range}
\end{figure}

However, the above format does not match the distribution of the target model, shown as the blue block area in Fig.~\ref{fig:weight-dynamic-range}. To avoid this mismatch, we change the bias to shift the distribution to the red dotted block area. Based on the results of the network simulation, we adopt a new bias 15 instead of the original bias 7 to cover most of the distributions of weights and activations. The dynamic range for the 8-bit floating-point is thus shifted from the blue block area to the red block area, as shown in Fig.~\ref{fig:weight-dynamic-range}. The quantization method used in this paper is the quantization aware training \cite{quant-aware-training} to quantize the weight and activation to different target bit numbers (4-, 8-, and 16-bit are tested in this paper).

\subsection{Model decomposition for depthwise dilated convolution and transposed convolution}
The depthwise dilated convolution in the separator network and the transposed convolution in the decoder network contain many zeros in the weights or inputs, resulting in a high computational redundancy. To reduce these redundancies, a 1-D model decomposition scheme instead of our previous 2-D decomposition work\cite{our_dt} is proposed to remove the computations with zero weights or activations for dilated and transposed convolutions in a structure way to ease the following hardware control.

\subsubsection{Dilated convolution}
\label{subsection:dilated-decompose}
\begin{figure}[t]
	\centering{\includegraphics[width=0.40\textwidth]{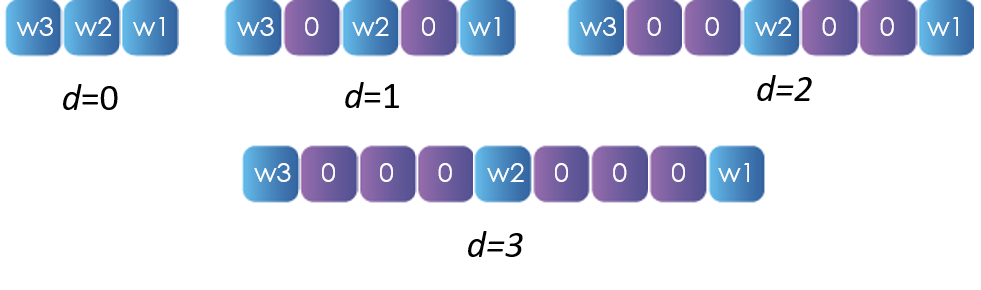}}
	\caption{The dilated weight with different dilation rate $d$}
	\label{fig:dilated}
\end{figure}

\begin{figure}[t]
	\centering{\includegraphics[width=0.38\textwidth]{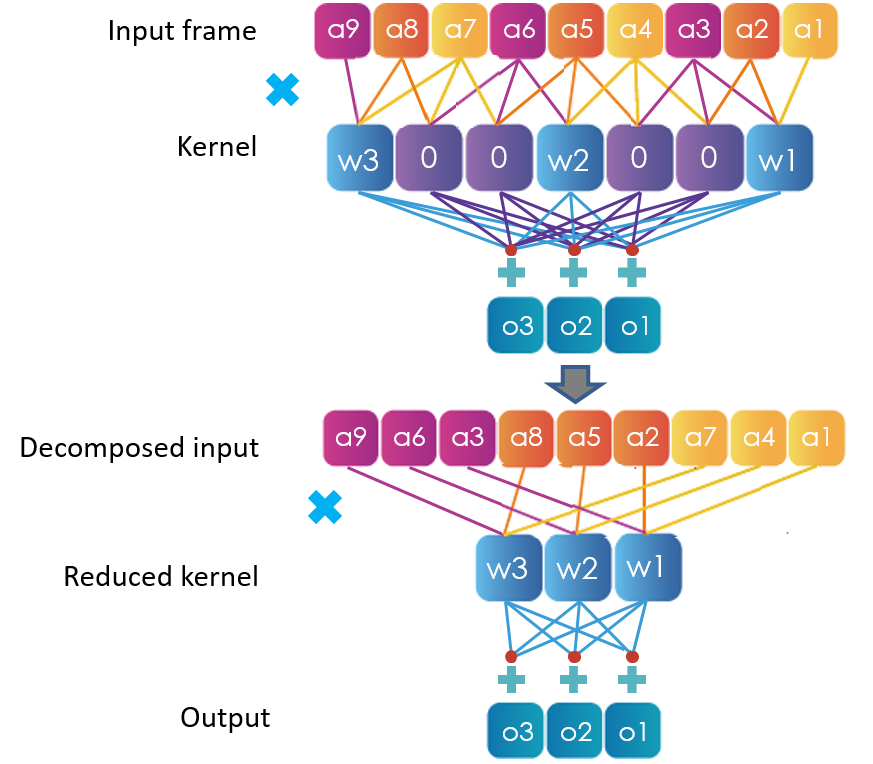}}
	\caption{The decomposition example for dilated convolution}
	\label{fig:dilated-decomposition}
\end{figure}

The dilated convolution performs the convolution operations using a kernel with zero insertions, as shown in Fig.~\ref{fig:dilated} with a kernel size of 3 and different dilation rate of $d$. Fig.~\ref{fig:dilated-decomposition} shows the dilated decomposition operation with a dilation rate of 2 as an example.  
Due to the dilation rate of 2, two zeros are inserted between the adjacent weights. Before applying the decomposition scheme, the outputs are calculated as 
$o_1 = a_7 \times w_3$ + $a_6 \times 0$ + $a_5 \times 0$ + $a_4 \times w_2$ + $a_3 \times 0$ + $a_2 \times 0$ + $a_1 \times w_1$;
$o_2 = a_8\times w_3$ + $a_7 \times 0$ + $a_6 \times 0$ + $a_5 \times w_2$ + $a_4 \times 0$ + $a_3 \times 0$ + $a_2 \times w_1$;
$o_3 = a_9\times w_3$ + $a_8 \times 0$ + $a_7 \times 0$ + $a_6 \times w_2$ + $a_5 \times 0$ + $a_4 \times 0$ + $a_3 \times w_1$.

To avoid computations with zeros, we decompose every $d+1$ input as a group and reorder the computation in a compact way. As shown in Fig.~\ref{fig:dilated-decomposition}, with the 1-D decomposition scheme, the input $a_1$-$a_9$ are decomposed into $a_1$$a_4$$a_7$, $a_2$$a_5$$a_8$ , and $a_3$$a_6$$a_9$ groups, while the weights are reduced to $w_1$$w_2$$w_3$ without zero insertions. 
Then, we can calculate the outputs as $o_1 = a_7\times w_3$ + $a_4 \times w_2$ + $a_1 \times w_1$; $o_2 = a_8\times w_3$ + $a_5 \times w_2$ + $a_2 \times w_1$; $o_3 = a_9\times w_3$ + $a_6 \times w_2$ + $a_3 \times w_1$; In this example, 57.1\% of computation complexity can be reduced by removing the zero computations using the proposed 1-D decomposition scheme.

\subsubsection{Transposed convolution}
\label{subsection:transposed-decompose}
\begin{figure}[t]
	\centering{\includegraphics[width=0.40\textwidth]{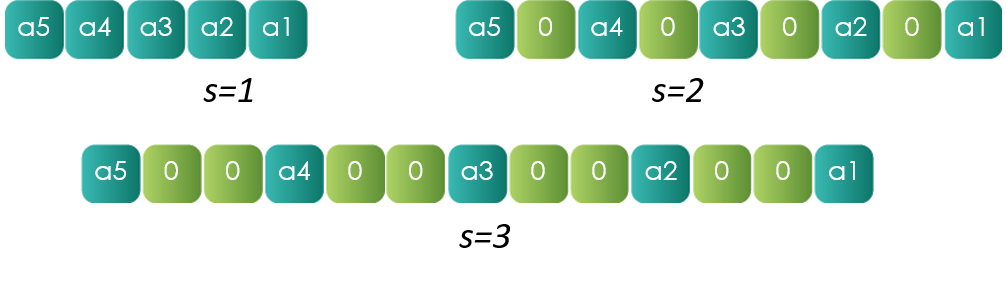}}
	\caption{The transposed inputs with different stride value $s$}
	\label{fig:transposed}
\end{figure}

\begin{figure}[t]
	\centering{\includegraphics[width=0.42\textwidth]{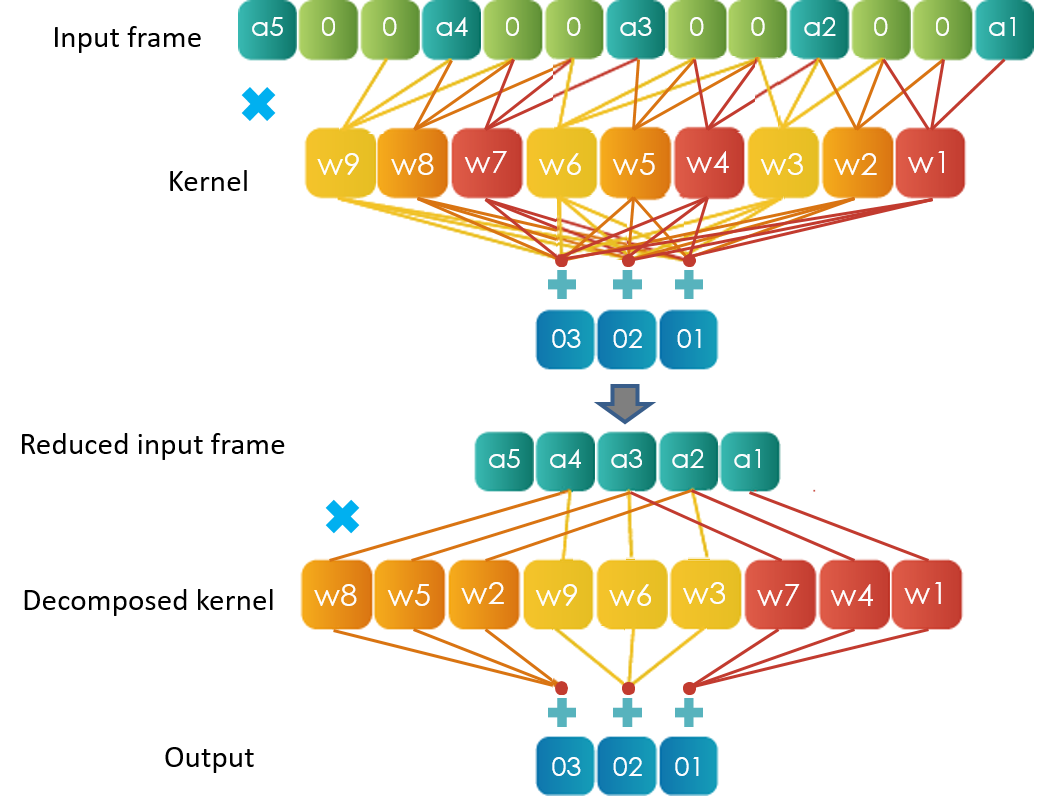}}
	\caption{The decomposition example for transposed convolution}
	\label{fig:transposed-decomposition}
\end{figure}

The transposed convolution in the decoder network inserts zeros between the adjacent inputs as shown in Fig.~\ref{fig:transposed} with input size of 5 and different stride value $s$, and then performs the convolution operation with the filter kernel. When taking the stride of 3 as an example, two zeros are inserted between the adjacent inputs. Fig.~\ref{fig:transposed-decomposition} shows the decomposition flow for transposed convolution. Before applying the decomposition scheme, the outputs are 
$o_1 = 0\times w_9$ + $0\times w_8$ + $a_3\times w_7$ + $0\times w_6$ + $0\times w_5$ + $a_2\times w_4$ + $0\times w_3$ + $0\times w_2$ + $a_1\times w_1$;
$o_2 = a_4\times w_9$ + $0 \times w_8$ + $0\times w_7$ + $a_3\times w_6$ + $0 \times w_5$ + $0 \times w_4$ + $a_2 \times w_3$ + $0 \times w_2$ + $0 \times w_1$;
$o_3 = 0\times w_9$ + $a_4 \times w_8$ + $0 \times w_7$ + $0 \times w_6$ + $a_3 \times w_5$ + $0 \times w_4$ + $0 \times w_3$ + $a_2 \times w_2$ + $0 \times w_1$.

After the decomposition process, the inputs are reduced to $a_1$-$a_5$ by removing the zeros between inputs, while the weights $w_1$-$w_9$ are decomposed into $w_1$$w_4$$w_7$, $w_3$$w_6$$w_9$, and $w_2$$w_5$$w_8$ groups. Then, the convolution outputs of $o_1$, $o_2$, $o_3$ are computed as $o_1=a_1 \times w_1$ + $a_2 \times w_4$ + $a_3 \times w_7$, $o_2= a_2 \times w_3$ + $a_3 \times w_6$ + $a_4 \times w_9$, and $o_3 =a_2 \times w_2$ + $a_3 \times w_5$ + $a_4 \times w_8$, every 3 cycles, respectively. By applying the proposed 1-D decomposition scheme, 66.7\% reduction of the computation complexity for the transposed convolution can be reduced by removing the zero operations of the multiplication and accumulation.

\begin{figure}[t]
	\centering{\includegraphics[width=0.45\textwidth]{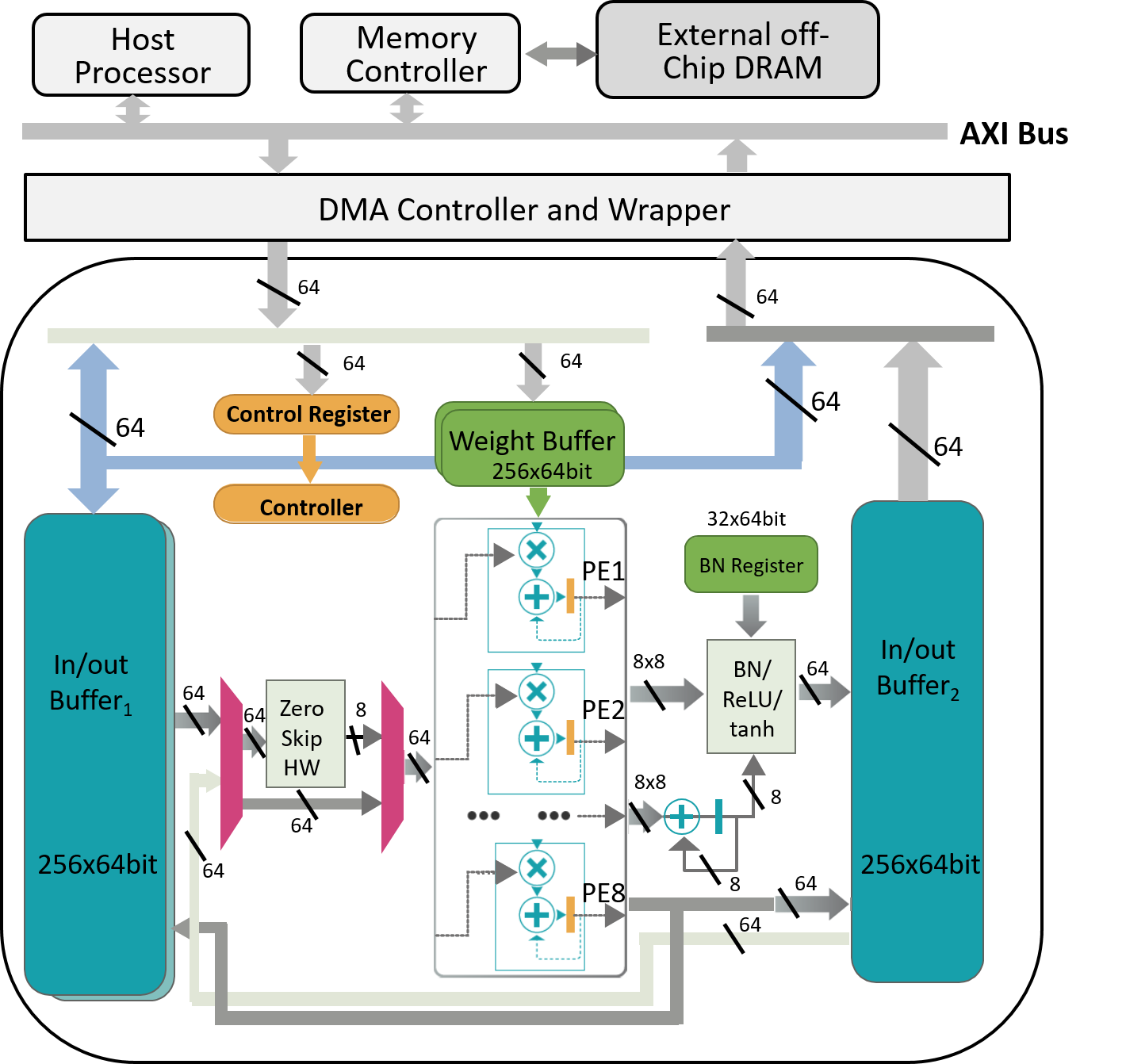}}
	\caption{The system architecture}
	\label{fig:system-architecture}
\end{figure}

\section{System Architecture}
\subsection{Overview}
The proposed deep learning accelerator aims at the real-time and low power execution for the speech separation application. For a 32 $ms$ frame of speech data with the sampling rate of 16 K samples/sec, the deep learning accelerator needs to process the speech computations within 32 $ms$ to guarantee the real-time execution. After applying the proposed model compression and decomposition techniques cooperated with the proposed zero skipping hardware, the total MAC operations of 34.848 M are needed to perform the speech separation operations for 32 $ms$ speech data. That means the computations of 1.089 GMACs/s for 16 KHz input signals is needed. Currently, the number of 8-bit floating-point processing elements in the accelerator is 8, two 256$\times$64-bit in/out buffer$_1$, two 256$\times$64-bit weight buffers, and a 256$\times$64-bit in/out buffer$_2$ are adopted in this accelerator. Since the model size and computation complexity of the speech separation model is greatly reduced by model compression and decomposition techniques as well as the proposed simple zero skipping hardware, the complexity of the accelerator hardware can thus be reduced to be very simple. When the DLA hardware operates at the frequency of 150 MHz, this accelerator hardware is capable of dealing with the real-time speech data and achieves the low power design with the simple hardware design.

Fig.~\ref{fig:system-architecture} shows the architecture of the proposed deep learning accelerator for speech separation application. This accelerator design consists of the 8 parallel 8-bit floating-point processing elements (PEs) for independent multiplication and accumulation operations, a zero skipping hardware, a unit for normalization and activation computation, two weight buffers, the in/out buffers, and a controller unit. In this architecture, we adopt the ping-pong buffer style for the in/out buffer$_1$ and weight buffer to enable concurrent read/write operations from processing element units and external DRAM. The in/out buffer$_1$ and buffer$_2$ also enable layer fusion operations and thus reduce the requirement of external memory bandwidth. The controller unit deals with the system flow control and generates the buffer addresses for the execution of the network model. The proposed deep learning accelerator communicates with a wrapper and a DMA controller, which connect to the ARM AXI bus to perform data movement, data format conversion and data rate adaptation. 
 
For model execution, the target multiresolution separation network consists of four main key operations as shown in Fig.~\ref{fig:tasnet}, including the 1-D convolution, the 1-D depthwise dilated convolution, the 1×1 pointwise convolution, and the 1-D transposed convolution. These data flows corresponding to these four main convolutional operations are described in Section IV-C to IV-F.

\subsection{Zero skipping hardware}
To reduce the computation complexity, a zero skipping hardware with broadcasting mechanism is proposed for sparse activation processing. As mentioned in Section~\ref{subsection:Pruning}, the structured sensitive pruning removes layers and channels, and the zero weight operations are skipped naturally. The unstructured pruning further prunes weights irregularly, but the sparsity ratio is low since most unimportant weights have been pruned by structured sensitive pruning. In contrast, the zero activations after the ReLU activation function introduce many zeros during the execution of separation model.  To simply the hardware design, the zero skipping hardware focuses on skipping the zero activation operations. All the PEs in DLA keep busy during the zero skipping operations, and the hardware can thus be fully utilized. The proposed zero skipping hardware resolves the problems of low hardware utilization and PE load imbalance in \cite{nctulaisparsecnn, dateloadbalance, cambricon-x}. Besides, to keep the DLA hardware efficient, our proposed zero skipping hardware performs the on-the-fly non-zero activation index detection, the additional index buffer used in \cite{cnvlutin, cambricon-x, nctulaisparsecnn, dateloadbalance, hansoneie} is not needed for each activation value. Thus, at least 12.5\% memory storage for the activation index can be saved due to the on-the-fly index detection scheme.  

\begin{figure}[t]
	\centering{\includegraphics[width=0.5\textwidth]{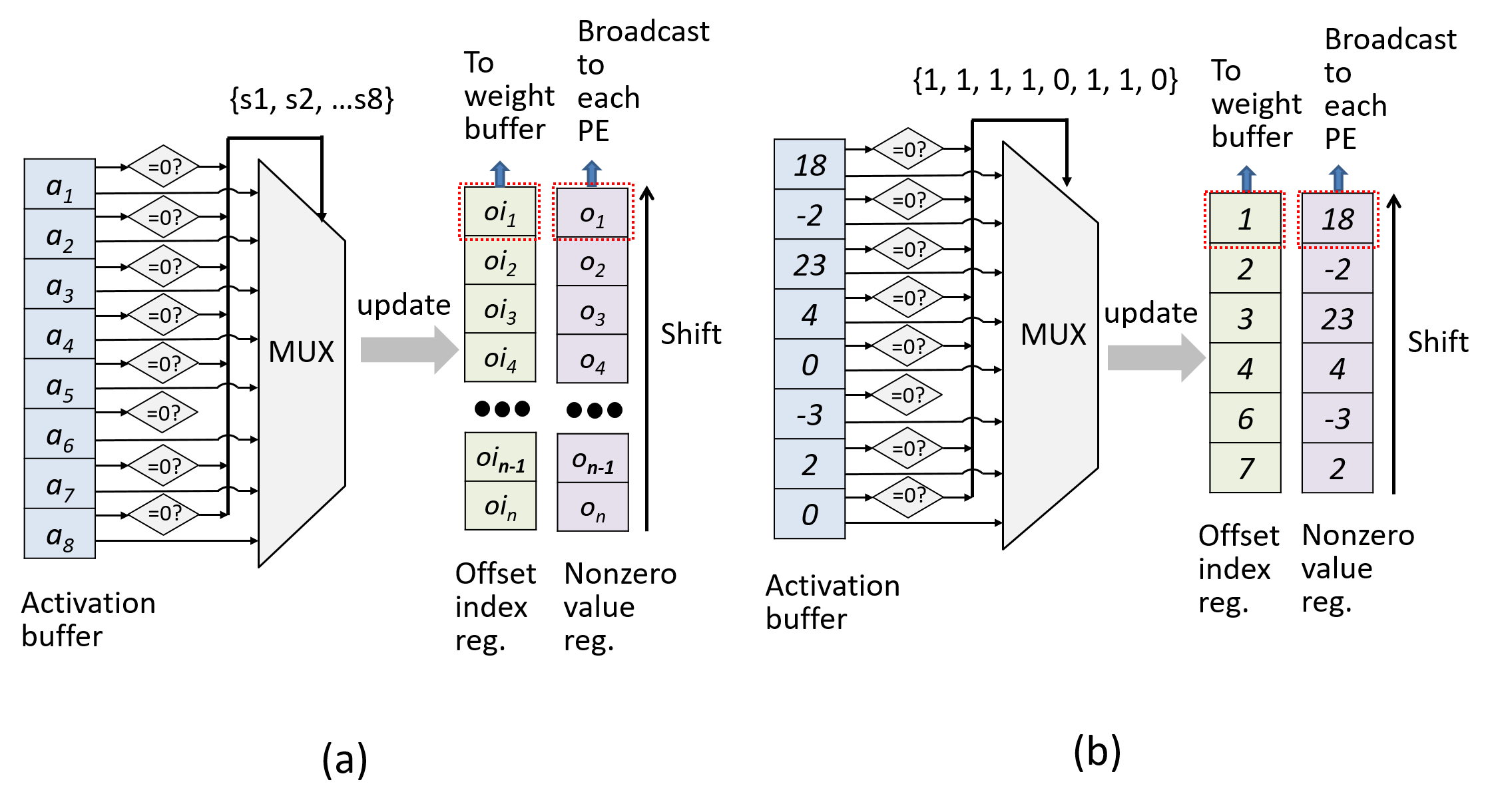}}
	\caption{Zero skipping hardware (a) Block diagram (b) An example }
	\label{fig:zero-skipping-hw}
\end{figure}

Fig.~\ref{fig:zero-skipping-hw}(a) shows the block diagram of our proposed zero skipping hardware. The zero skipping hardware is very simple due to the proposed simple 8-parallel MAC architecture. The eight 8-bit activation data are first fetched from the in/out activation buffer, and then the eight zero-comparators are used to decide eight 1-bit nonzero activation indexes on-the-fly. These eight 1-bit indexes are then concatenated as \{s1, s2, ..., s8\} and sent to the selection signals of the nonzero value multiplexer. The multiplexer selects and decides the offset indexes and nonzero activation values accordingly and then stores them in the corresponding registers. One offset index and one 8-bit nonzero activation value are then shifted out in order every cycle, and sent to weight address controller for address generation and broadcast to the input of eight PEs, respectively, until all the offset indexes and nonzero activation values are shifted out. Fig.~\ref{fig:zero-skipping-hw}(b) shows an example for the zero skipping operations. The comparators decide eight 1-bit nonzero indexes as \{1, 1, 1, 1, 0, 1, 1, 0\}. These eight indexes are used as the selection signals in the multiplexer and then the six nonzero offset indexes and values \{1, 18\}, \{2, -2\}, \{3, 23\}, \{4, 4\}, \{6, -3\} and \{7, 2\} are selected on-the-fly and stored in the offset index and nonzero value registers. These six nonzero offset indexes and values are shifted out from the registers in order every cycle and sent to the weight address controller and broadcast to the inputs of eight PEs, respectively, for the convolution operations.

\subsection{Data flow for the 1-D convolution}
\begin{figure}[t]
	\centering{\includegraphics[width=0.45\textwidth]{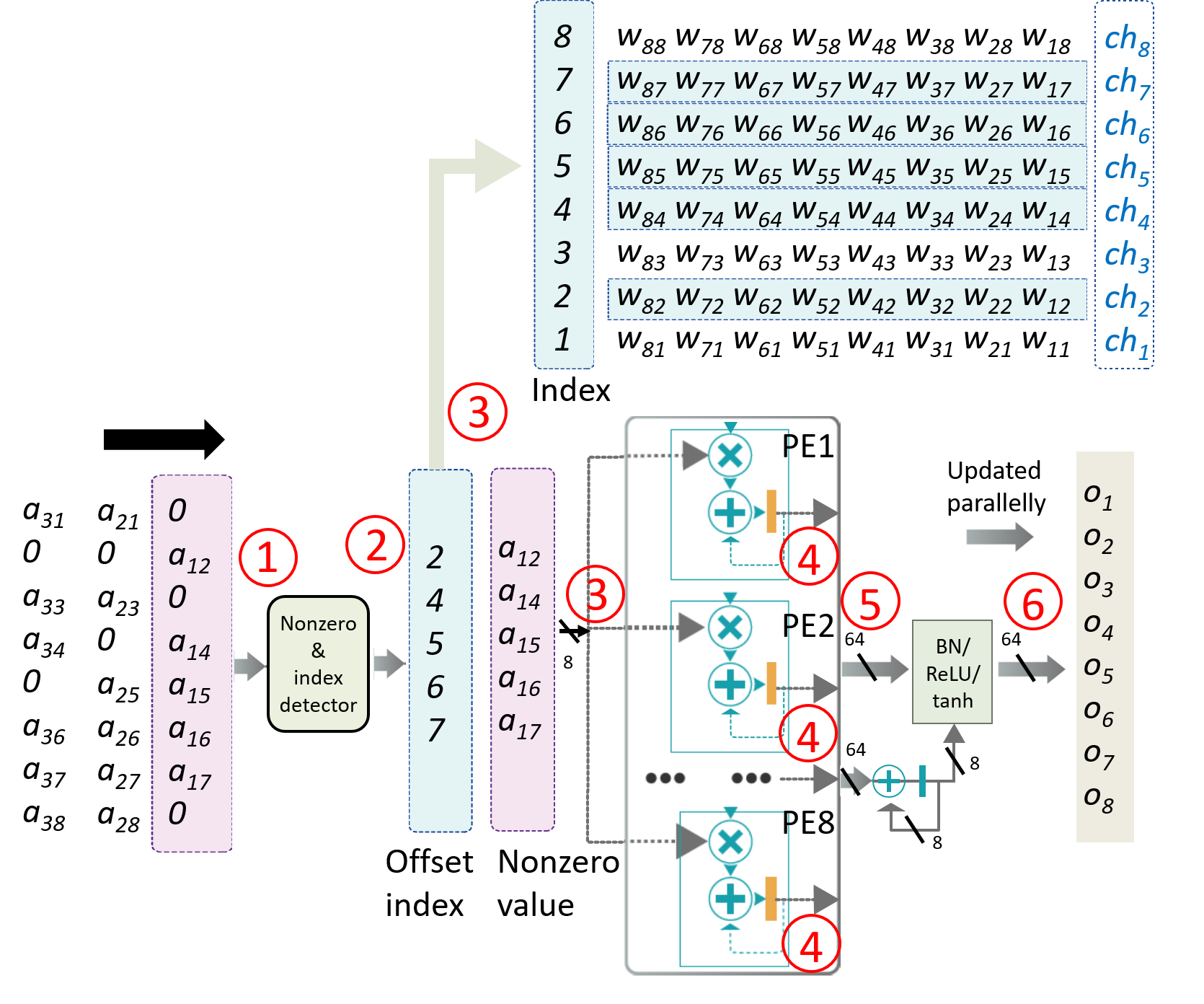}}
	\caption{Data flow for 1-D convolution}
	\label{fig:dla-encoder}
\end{figure}

Fig.~\ref{fig:dla-encoder} shows the data flow of the 1-D convolution that consists of six stages. In stage 1, a 64-bit input from the in/out buffer$_1$ or the in/out buffer$_2$ is selected and sent to the nonzero/index detector, the detector then decides and stores the nonzero offset index and value into the registers in stage 2.  In stage 3, the nonzero offset index is shifted out and added with the base address of the weight buffer and the corresponding eight 8-bits weights are read. On the other side, an 8-bit nonzero activation value is shifted out from the nonzero value register, and then broadcast to each PE. In stage 4, each PE multiplies the nonzero activation with its weight from one channel every cycle and accumulates the previous multiplication result until the channel operations are completed. In this way, the input data is shared by broadcasting to all PEs and the PE output values are reused locally until the computations are completed. Take PE$1$ as an example, the PE$1$ output will be computed as $a_{12}\times w_{12}$ + $a_{14}\times w_{14}$ + $a_{15}\times w_{15}$ + $a_{16}\times w_{16}$ + $a_{17} \times w_{17}$ for the input sequence $a_{12}\  a_{14}\ a_{15}\ a_{16}\ a_{17}$. In stage 5, the 8 PE results are sent to the normalization and activation units, and then saved to the eight 8-bit registers in parallel and written to the in/out buffer$_2$ in stage 6.

\begin{figure}[t]
	\centering{\includegraphics[width=0.48\textwidth]{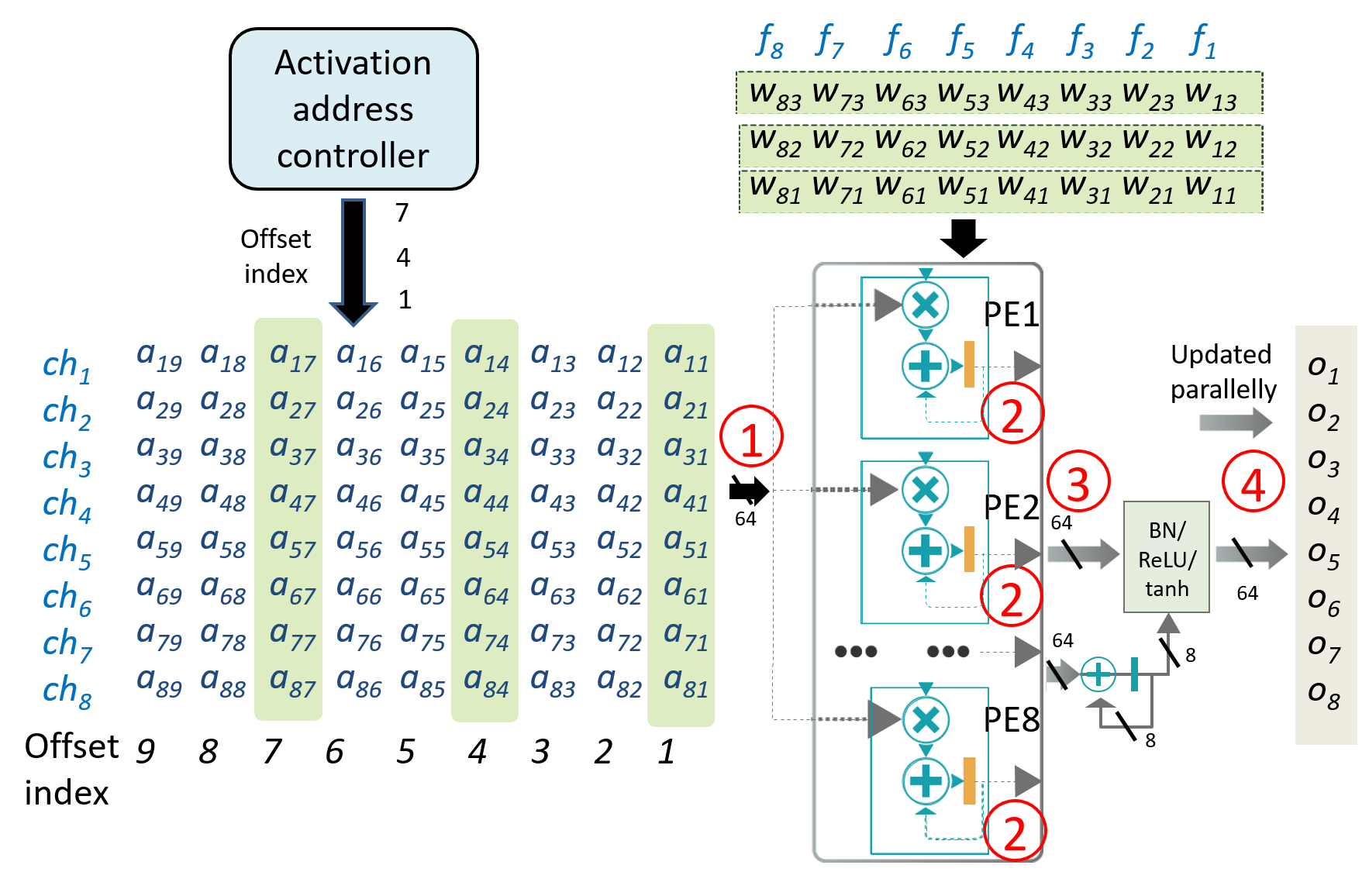}}
	\caption{Data flow for dilated convolution in DLA}
	\label{fig:dla_dilated}
\end{figure}

\subsection{Data flow of the depthwise dilated convolution}
The second key operation in the speech separation network is the 1-D depthwise dilated convolution. By using the proposed 1-D decomposition scheme mentioned in Section~\ref{subsection:dilated-decompose}, the inserted zeros in adjutant weights are removed naturally, while the inputs are decomposed via the activation address controller.  Fig.~\ref{fig:dla_dilated} shows the data flow for the depthwise dilated convolution for the case of a weight size of 3 and a dilated value of 2. Since the input data for each channel in the depthwise convolution is not reused by other filters, only the nonzero index but the broadcasting mechanism in zero skipping hardware is used to generate the gated clock signals for the corresponding PEs. In stage 1, we select eight 8-bit inputs belonging to eight channels according to the offset index generated by the activation address controller from in/out buffer$_1$ or buffer$_2$ to eight PEs, respectively. Each processing element performs the MAC operation every cycle until all the channel computations are completed in stage 2. In this stage, the processing element performs the output stationary due to the output data reuse in each MAC iteration. The channel outputs of eight processing elements are sent to the normalization and activation unit in stage 3, and then the final outputs are sent to the in/out buffer$_1$ or buffer$_2$ in parallel in stage 4. Take the weight size of 3 and dilated value of 2 as an example. The eight 8-bit decomposed inputs are fetched according to the offset index generated by the activation address controller, and the corresponding zero-removed weights are also read in stage 1 as shown in Fig.~\ref{fig:dla_dilated}. The output of PE$1$ can be calculated as $o_1 = a_{11} \times w_{11}$ + $a_{14} \times w_{12}$ + $a_{17} \times w_{13}$. The outputs of PE$2$-PE$8$ can be calculated in the same way. After the outputs of PE$1$-PE$8$ performs the normalization and activation operations, the outputs $o_{1}$-$o_{8}$ are obtained and stored in eight 8-bit registers in parallel and written to the in/out buffer$_2$.

\subsection{Data flow for 1×1 pointwise convolution}
\begin{figure}[t]
	\centering{\includegraphics[width=0.47\textwidth]{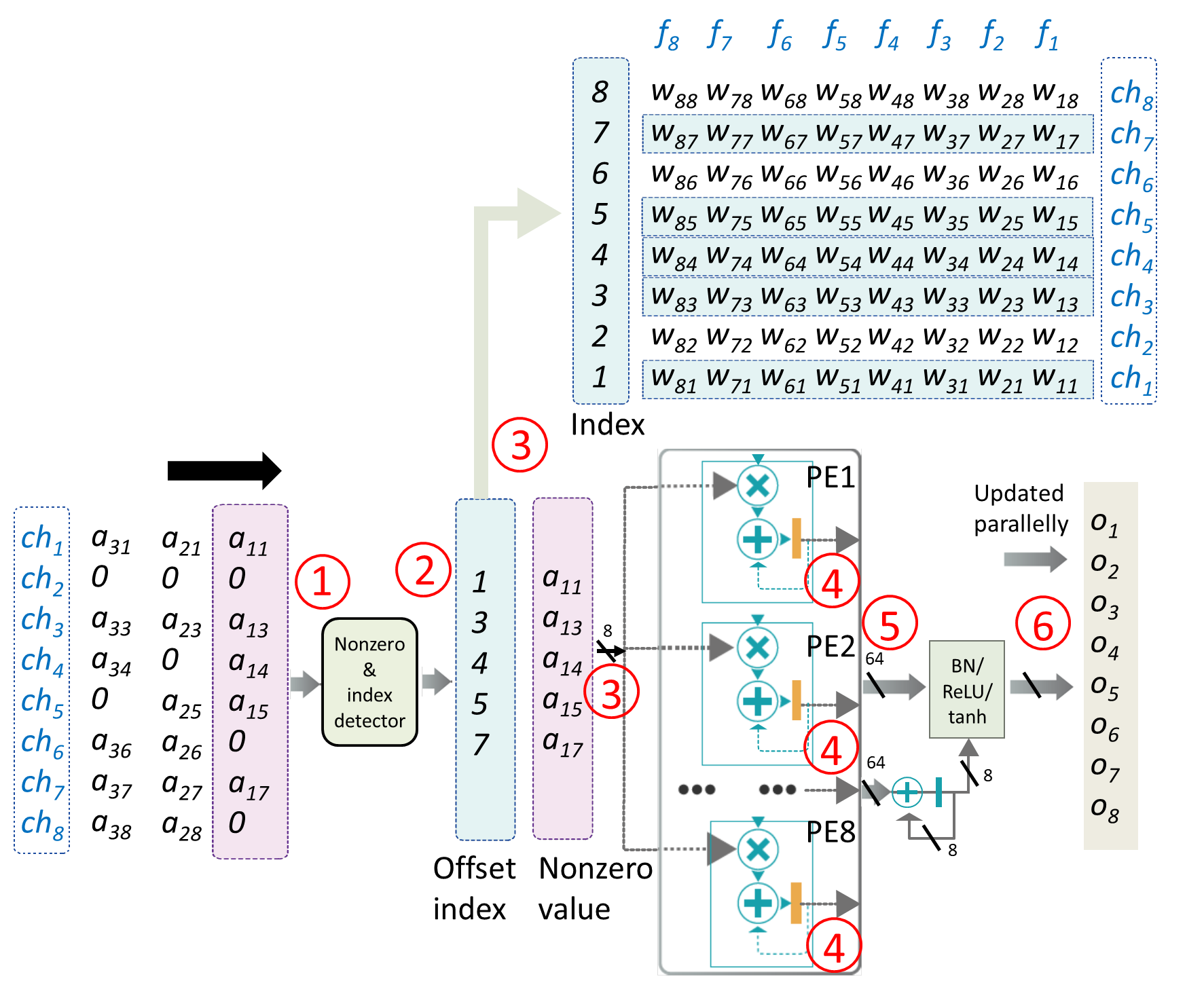}}
	\caption{Data flow for 1×1 pointwise convolution in DLA}
	\label{fig:dla-1x1}
\end{figure}

The 1×1 pointwise convolution performs the operations for all channels of each point, and this kind of operation is often used to change the number of the output channels. Fig.~\ref{fig:dla-1x1} shows the 1×1 pointwise convolution data flow when executing this operation in our 8-parallel MAC accelerator hardware. The 1x1 pointwise operation flow in DLA is similar to the 1-D convolution data flow. A 64-bit input from the in/out buffer$_1$ or the in/out buffer$_2$ is selected and sent to the nonzero/index detector, and the detector then decides and stores the nonzero offset indexes and values into the registers in stage 1 and 2. 
One nonzero offset index is shifted to obtain the corresponding eight 8-bit weights and then the weights are sent to corresponding PEs. One 8-bit nonzero activation value is also shifted to broadcast to each PE in stage 3. Each PE multiplies with a weight from one channel every cycle and accumulates previous multiplication results until all channel operations are completed in stage 4. In this way, the input data is shared and broadcast to all PEs and the PE output values are reused locally until the computations are completed. The output is sent to the normalization and activation unit in stage 5, and the final outputs are sent to the output registers and in/out buffer$_2$ in stage 6. Take the PE$2$ as an example, the output of PE$2$ is calculated as $a_{11} \times w_{21}$ + $a_{13} \times w_{23}$ + $a_{14} \times w_{24}$ + $a_{15} \times w_{25}$ + $a_{17} \times w_{27}$. The outputs of PE$1$ and PE$3$-PE$8$ are calculated in the same way. The PE outputs are sent to the normalization and activation unit, and then the eight outputs are obtained and stored in eight 8-bit registers in parallel and written to the in/out buffer$_2$.  

\begin{figure}[t]
	\centering{\includegraphics[width=0.39\textwidth]{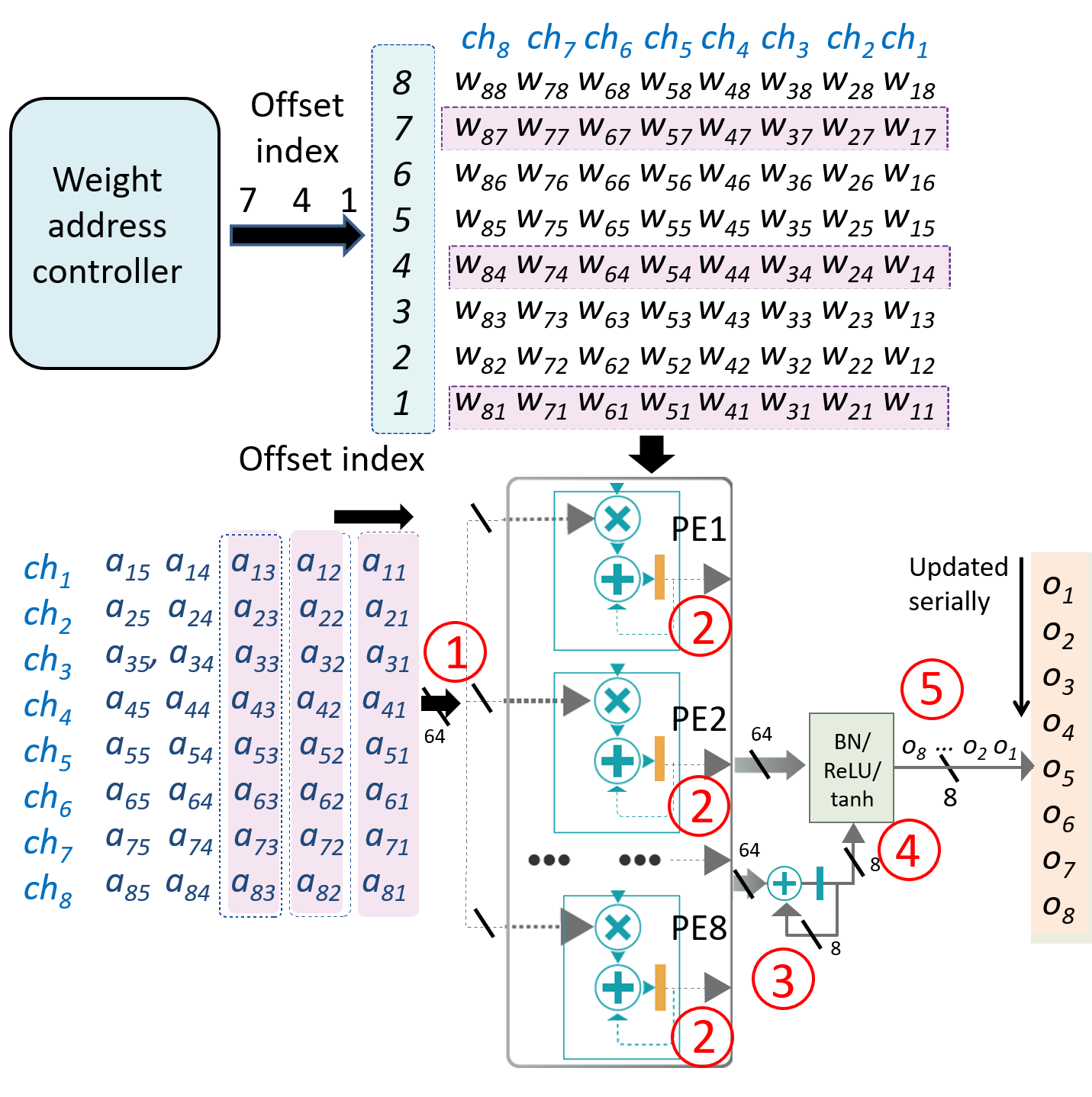}}
	\caption{Data flow for 1-D transposed  convolution in DLA}
	\label{fig:dla-decoder}
\end{figure}

\subsection{Data flow for 1-D transposed convolution}
The decoder performs the transposed convolutions by using a short- and a long-length 1-D filters to reconstruct two separated signals from the separator network. By using the proposed 1-D decomposition scheme in Section~\ref{subsection:transposed-decompose}, the inserted zeros between adjacent inputs are removed naturally while the filter kernel is decomposed through the weight address controller. In the case of stride size of 3, Fig.~\ref{fig:dla-decoder} shows the 1-D transposed convolution data flow when applying this operation to our 8-parallel MAC units. Since the number of separation output channels is 2, to keep the DLA busy, the input data for each channel in the transposed convolution is not reused by the other filters. That means the input broadcasting mechanism is not used to share the inputs, and only the nonzero index detection is used to generate the gated clock signals to the corresponding PEs for low power processing. The input data of each different channel is fed to different processing elements, while the decomposed weight data is also fed according to the offset index value generated by the weight address controller to processing elements for the MAC operations in stage 1 and 2. The operations in stage 2 are repeated until all the channel computations are completed. The outputs of the 8 processing elements are summed together in stage 3 and sent to the normalization and activation unit in stage 4. The final outputs are sent to output registers in serial and then written to in/out buffer$_2$ in stage 5. When taking the stride value of 3 as an example, the sequences of the zero-removed inputs and decomposed weights are shown in Fig.~\ref{fig:dla-decoder}. The output sequence of PE$1$ can be calculated in order as $a_{11} \times w_{11}$ + $a_{12} \times w_{14}$ + $a_{13} \times w_{17}$. The outputs of PE$2$-PE$8$ can be calculated in the same way. The outputs of PE$1$-PE$8$ are summed together in the accumulator, and then sent to the normalization and activation unit. The eight outputs of $o_{1}$-$o_{8}$ are computed and obtained serially and updated to eight 8-bit registers in order and then sent to the in/out buffer$_2$.

\section{Experimental Results}

The proposed model is evaluated on the DSD100 dataset \cite{dsd100}. Its training data and the test data with 16K samples/sec sampling rate contain the mixed data of vocal and musical with the lengths of 57,310 secs and 8,820 secs, respectively. The model is optimized with the Adam optimizer and the learning rate set as $3e^{-4}$. The loss function $L$ is set as ${0.5}\times(vocal\ loss + musical\ loss)$. This paper adopts a multiresolution speech separation model \cite{chi_separation, chi_melody} as our baseline for comparisons with compressed models.

\begin{figure}[t]
	\centering{\includegraphics[width=0.45\textwidth]{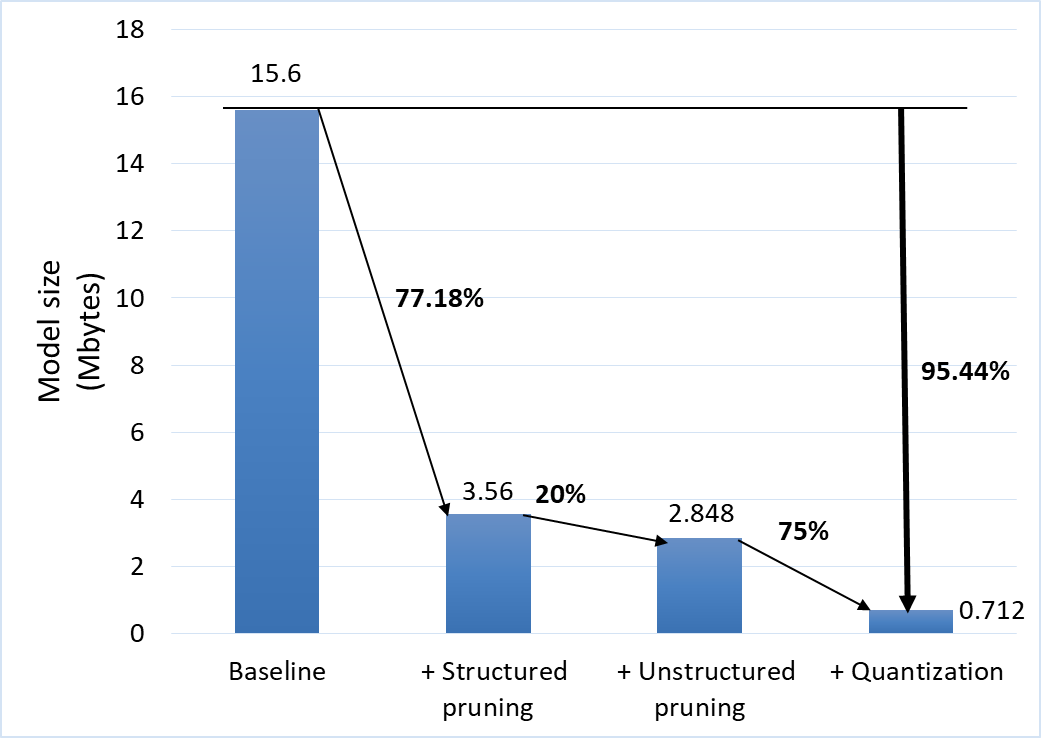}}
	\caption{The weight size reduction by different model compression techniques}
	\label{fig:weight-reduction}
\end{figure}

\begin{figure}[t]
	\centering{\includegraphics[width=0.48\textwidth]{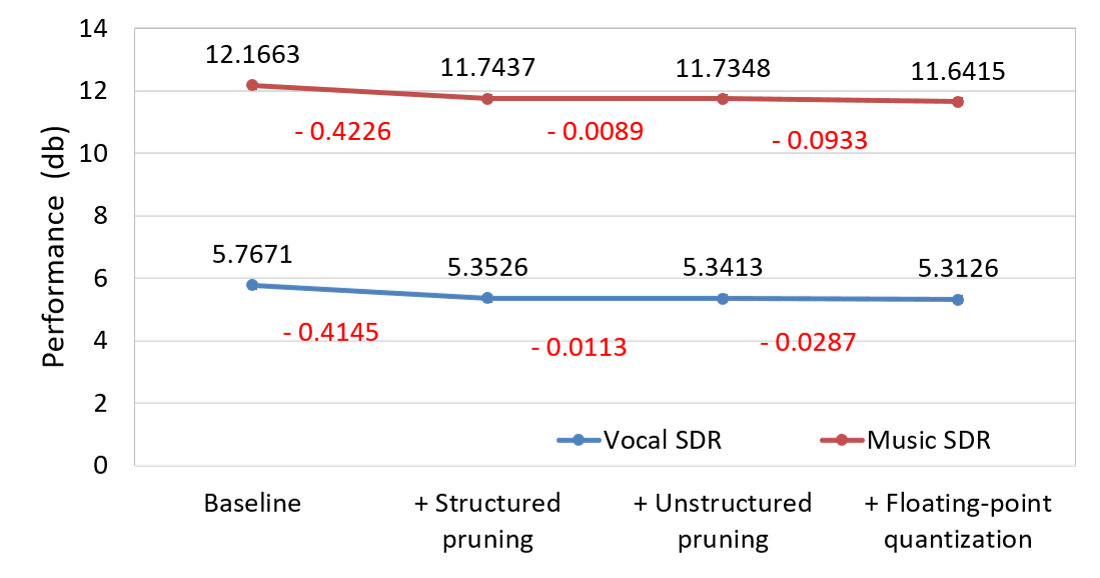}}
	\caption{The vocal and musical test accuracy using different model compression techniques}
	\label{fig:test-accuracy}
\end{figure}

\subsection{Model pruning and decomposition}

\begin{figure}[t]
	\centering{\includegraphics[width=0.45\textwidth]{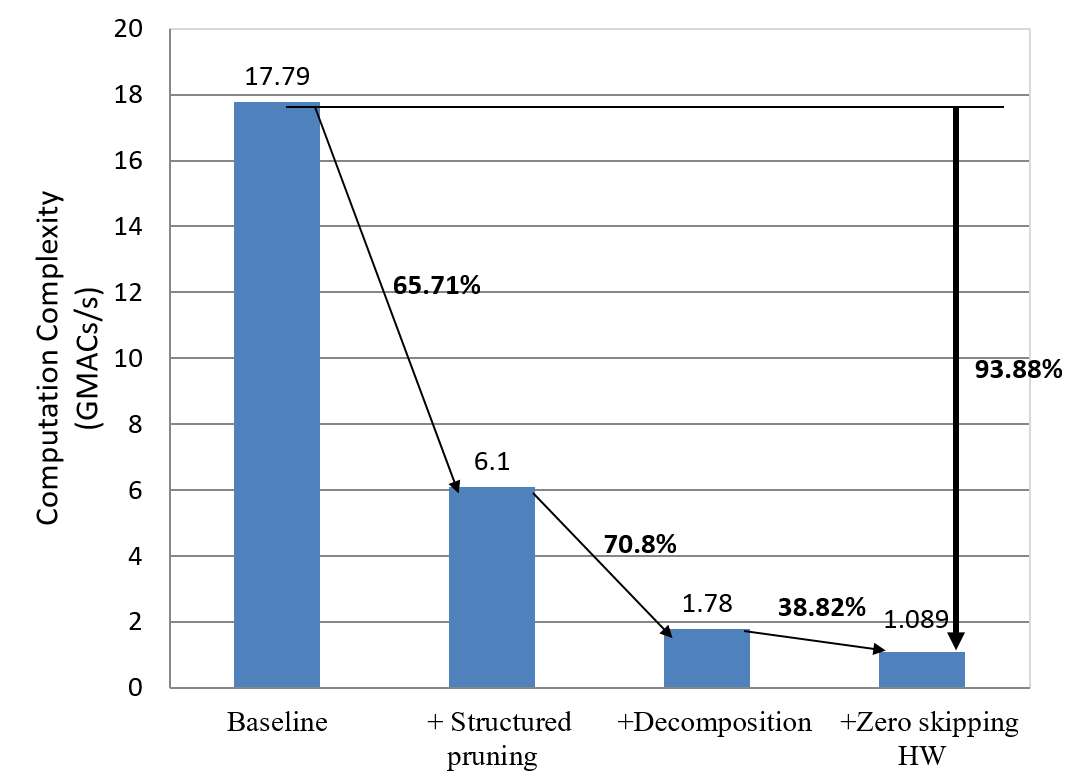}}
	\caption{The computational complexity with different techniques }
	\label{fig:computation-comparison}
\end{figure}

Fig.~\ref{fig:weight-reduction} to \ref{fig:computation-comparison} show the reduction in model size, the test accuracy and complexity using different compression techniques. The baseline model needs a model size of 15.6 MB as in Fig.~\ref{fig:weight-reduction}. After structured sensitivity pruning and unstructured pruning, the model size becomes 3.56 MB and 2.848 MB, that is, 77.18\% and 20\% weight reductions, respectively. For computational complexity, Fig.~\ref{fig:computation-comparison} shows that the number of MAC operations can be reduced to 6.1 GMACs/s and 1.78 GMACs/s, that is, a reduction of 65.71\% and 70.8\%, respectively, after the structured sensitivity pruning and decomposition of the model. The test precision has 0.4226 db of musical SDR and 0.4145 db of vocal SDR drop, respectively, for structured sensitivity pruning and almost stays the same for unstructured pruning, as shown in Fig.~\ref{fig:test-accuracy}.

\begin{figure}[t]
	\centering{\includegraphics[width=0.48\textwidth]{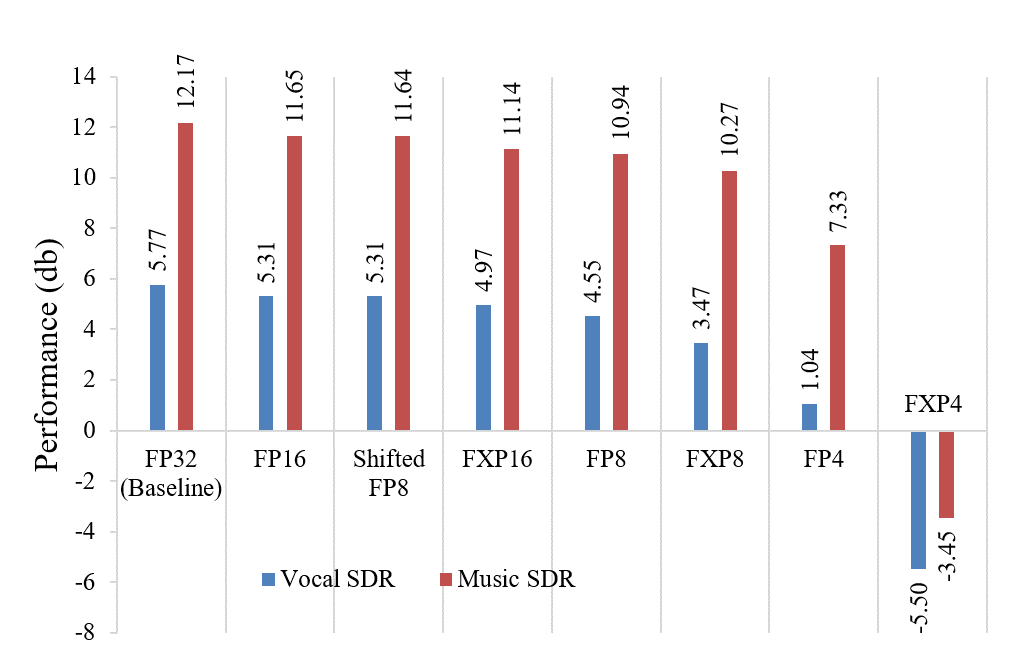}}
	\caption{The vocal and musical accuracy for different quantization data format and size based on the pruned network with the parameter of \{2, 256, 8, 3, 128\}.}
	\label{fig:quant-result}
\end{figure}

\subsection{Model quantization}

Fig.~\ref{fig:quant-result} shows the simulation results in fixed-point and floating-point formats of 4-, 8-, and 16-bit on the pruned network. The 4-bit format is too small to have good accuracy for speech applications compared to the 32-bit floating-point format. The 16- and 8-bit fixed-point and 8-bit floating-point formats have lower performance compared to the proposed 8-bit shifted floating-point format.  The 16-bit floating-point and shifted 8-bit floating-point formats have kept almost the same accuracy as its 32-bit counterpart, but the 16-bit floating-point format costs much more hardware overhead and memory storage. The shifted 8-bit floating-point quantization also saves 75\% of model size. The proposed format can provide a higher dynamic range to fit speech signals, but has a smaller area cost (129 ${\mu}m^2$) compared to the 8-bit fixed-point format (243 ${\mu}m^2$) for a multiplier design. This smaller area is because our proposed shifted 8-bit floating-point format only uses a 3-bit multiplier for mantissa multiplication, instead of using an 8-bit multiplier in the fixed-point format.

\begin{figure}[t]
	\centering{\includegraphics[height=42mm]{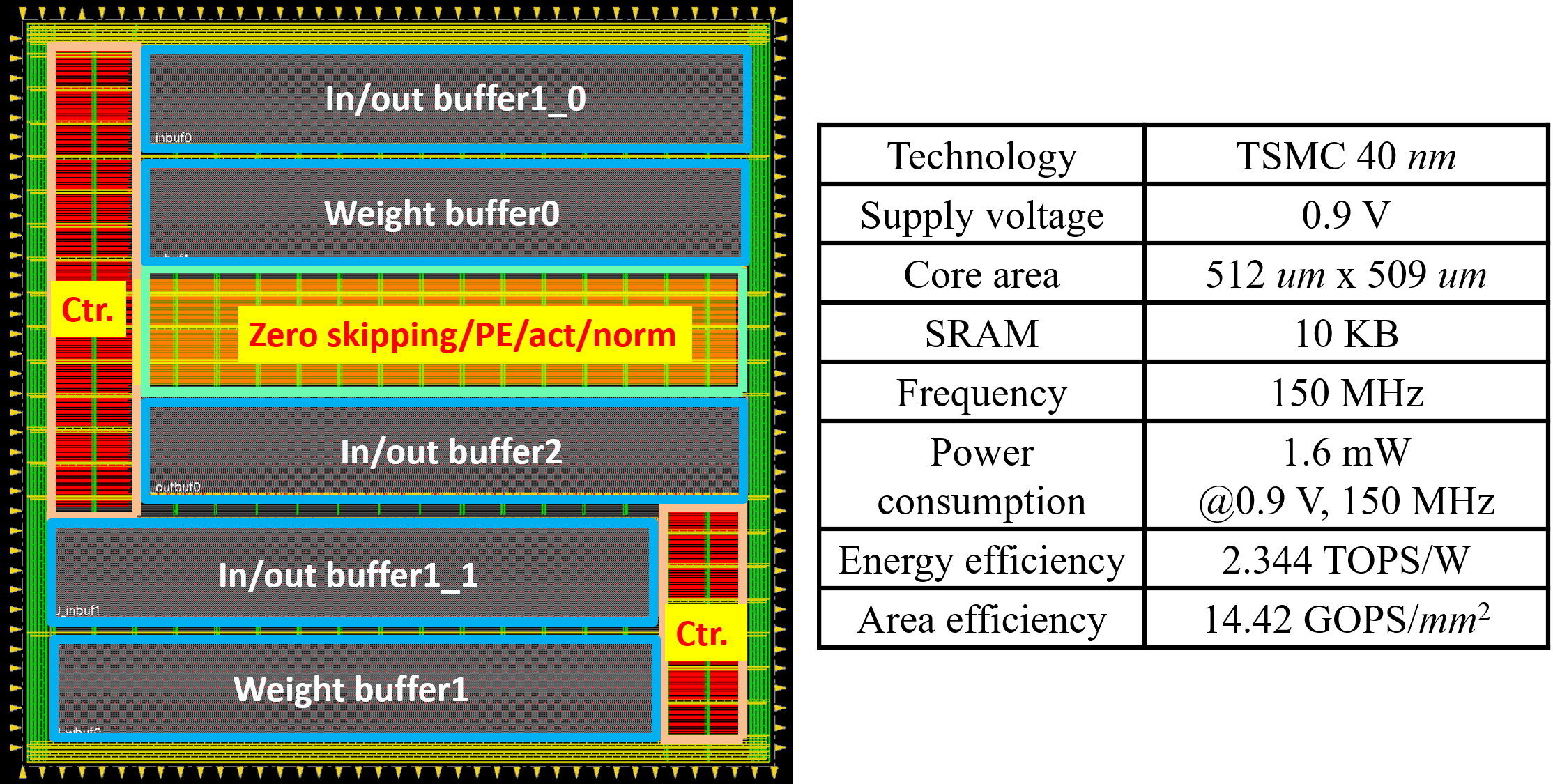}}
	\caption{The DLA design layout and performance summary}
	\label{fig:new_chip_photo}
\end{figure}

\begin{figure}[t]
	\centering{\includegraphics[width=0.52\textwidth]{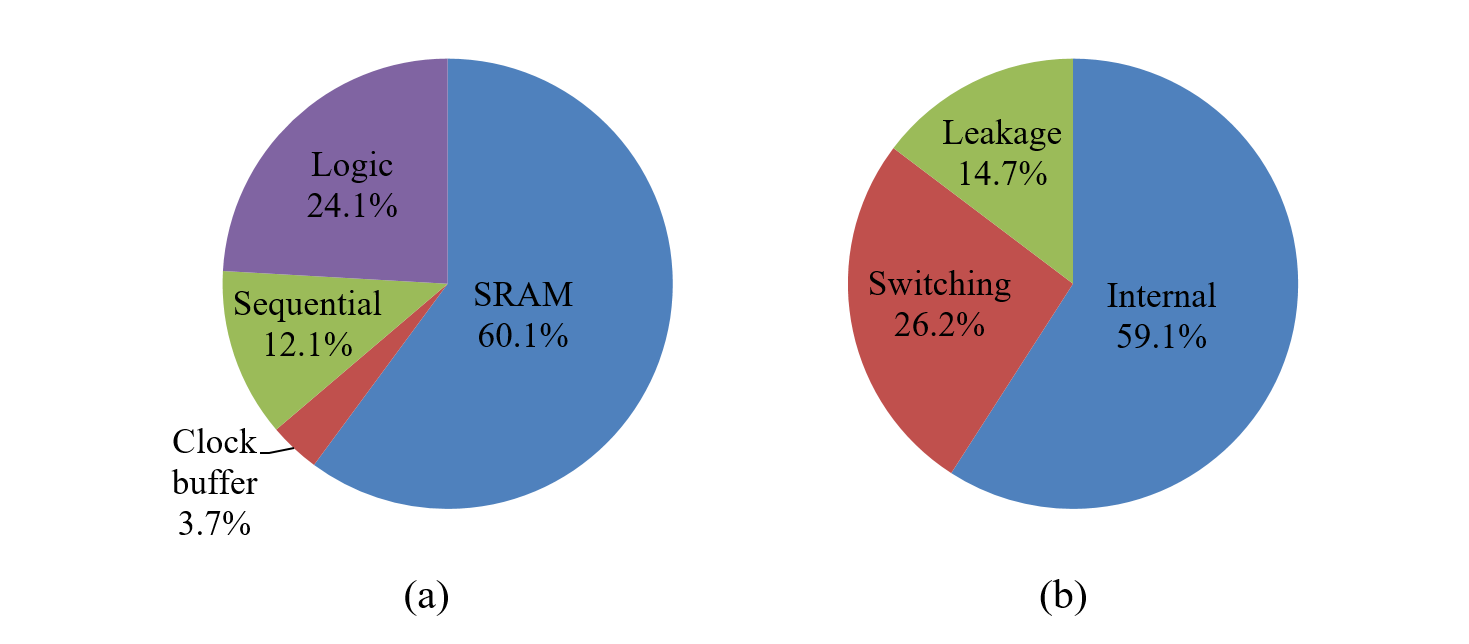}}
	\caption{The DLA design power breakdown}
	\label{fig:chip-power-breakdown}
\end{figure}

\subsection{Hardware implementation results}

Fig.~\ref{fig:new_chip_photo} shows the DLA design layout and the summary of performance. This accelerator design is implemented with the TSMC 40 $nm$ CMOS process. The core area is $512\ um \times 509\ um$ with 10 KB SRAM,  where the weight buffer, in/out buffer, zero skipping/PE/act/norm, and controller occupy about 21.47\%, 32.20\%, 11.08\%, and 10.66\%, respectively, based on the layout area. The total latency when executing the compressed speech separation model with 32 $ms$ speech data is 4.391311 M clock cycles. When the accelerator operates at the frequency of 150 MHz, the total processing time is 29.275 $ms$. That means the proposed low power accelerator is capable of executing the speech separation job in real-time. According to the results of post-layout simulation and power analysis, the design for the real-time execution consumes 1.6 mW power at the 150 MHz operating frequency and a supply voltage of 0.9 V, achieving a 2.344 TOPS/W energy efficiency and a 14.42 GOPS/$mm^2$ area efficiency, where TOPS and GOPS indicate Tera Operations Per Second and Giga Operations Per Second, respectively. Note that a higher clock rate is possible due to the simple architecture design. We attain this low power consumption through algorithm and architecture optimizations, as well as gating idle logic and buffers. 

Fig.~\ref{fig:chip-power-breakdown} shows the power breakdown of the implementation. The SRAM buffer consumes the highest power consumption, 60.1\%, while logic circuits, sequential circuits, and clock buffer consume the power consumption of 24.1\%, 12.1\% and 3.7\%, respectively. Internal power, switching power, and leakage power occupy 59.1\%, 26.2\%, and 14.7\% of total power consumption, respectively.  

\begin{table*}[t]
	\centering
	\caption{Comparisons with other DLA designs}
	\label{table:overview}
	\begin{tabular}{|l||c|c|c|c|c|c|c|}
\hline
\multicolumn{1}{|c|}{}  &  
\multicolumn{1}{c|}{JSSCC 2020 \cite{NTUADSP}} & 
\multicolumn{1}{c|}{TCAS-I 2020 \cite{NTUASP-2}}& 
\multicolumn{1}{c|}{{$^{d}$}ISSCC 2021 \cite{2021issccasr}} &
\multicolumn{1}{c|}{This Work} \\ \hline 

Technology       & 40\ nm      & 40\ nm  & 16\ nm  & 40\ nm     \\ \hline
Supply voltage    & 0.6 V  & 0.7 V & 0.55 V  & 0.9 V    \\ \hline
Application       & Hearing Devices &   Hearing Devices & Speech Recognition    & Speech  Separation      \\ \hline
Algorithm  & FFT+CNN+FC  & FFT+SCE & FFT+RNN  & CNN     \\ \hline
Dataset  & NOISEX-92  & Corpus & LibriSpeech  & DSD100     \\ \hline
Precision  & {FXP 8-bit}  & FXP 6-bit & FP 8-bit & Shifted FP 8-bit      \\ \hline
No. of NN layers  & 4  & No  & N/A  & 20     \\ \hline
No. of PEs  & 64  & N/A  & 4 & 8     \\ \hline
Sparse processing     & No         & No        & No   & Yes (Activation)             \\ \hline
Frequency (MHz)    & 5         & 10.5   & 573       & 150               \\ \hline
SRAM     & 327 KB           &  8 KB   & 5.03 MB      & 10 KB   \\ \hline
Core area  (mm$^2$)        & 4.2         & {$^{c}$}0.3   & 8.84    & 0.26    \\ \hline
Core power (mW)       & 2.17    & 1.5  & 19       & 1.6           \\ \hline

\multirow{2}{4cm}{Core energy efficiency (TOPS/W)}  &1.2  & N/A & 7.8  & 2.344 \\
  &{$^b$}0.53    & N/A & {$^b$}1.17  &  2.344    \\ \hline
  
\multirow{2}{4cm}{Core area efficiency (GOPS/mm$^2$)}  & 0.62   &  N/A  & 16.76    & 14.42  \\
  &{$^{a,b}$}0.27   &  N/A  & {$^{a,b}$}6.71    & 14.42    \\ \hline
  
		\multicolumn{1}	{|l}{$^{a}$Technology scaling ($\dfrac{process}{40\ nm}$).} & \multicolumn{4}	{l|}{$^{b}$Normalized~energy~efficiency $=$ energy~efficiency $\times(\dfrac{process}{40\ nm})\times(\dfrac{voltage}{0.9\ V})^2$.} \\
		\multicolumn{1}	{|l}{$^{c}$Logic area only.} & \multicolumn{4}	{l|}{{$^{d}$}FlexASR only.} \\

		\hline	
		
\end{tabular}
\end{table*}

\subsection{Design comparisons}
Comparisons to other designs are difficult and are not fair due to different applications and models. For a reference purpose, we select the designs for speech enhancement or speech related applications as listed in Table~\ref{table:overview}.

The designs in \cite{NTUADSP, NTUASP-2} are programmable acoustic signal processors for hearing aids devices, which use the conventional speech separation operations in the frequency domain. The design in \cite{2021issccasr} is an SoC design on frequency domain data for speech recognition application. In contrast, the proposed design supports four main  convolutional computations in time domain instead of frequency-domain as in \cite{NTUADSP, NTUASP-2, 2021issccasr}, which has greatly simplified the overall task. 

For design efficiency, our design has the best energy and area efficiency.  The lower power consumption and smaller area in \cite{NTUASP-2} are because it adopts the conventional speech algorithm rather than deep learning methods in other designs, which will have lower performance. Our better efficiency is due to the algorithm and architecture co-optimization.

\section{Conclusions}
This paper proposes a low power inference accelerator design for a fully time domain speech separation application. The proposed model compression scheme reduces the network model size as well as the computation complexity required by hardware. The decomposition and shifted floating-point quantization techniques are used to simplify the hardware design. A simple zero-skipping hardware cooperated with an 8 parallel independent MAC architecture for the sparse activation processing is adopted to further reduce the computation complexity while the PE hardware is fully utilized. Based on the above techniques, the overall network model complexity is greatly reduced, therefore, an 8 independent MAC architecture is sufficient to execute different 1-D convolution operations in the separation model. The final design is implemented with the TSMC 40 $nm$ process. A better normalized energy efficiency and area efficiency are obtained than other DLA designs. Deploying other speech-related applications to this DLA design is our next step to take in the future.

\section*{Acknowledgment}
The authors would like to thank Prof. Tai-Shih Chi for providing the baseline model, and TSRI for supporting EDA design tools.

\bibliographystyle{IEEEtran}
\bibliography{IEEEabrv, thesis}

\begin{IEEEbiography}[{\includegraphics[width=1in,height=1.25in,clip,keepaspectratio]{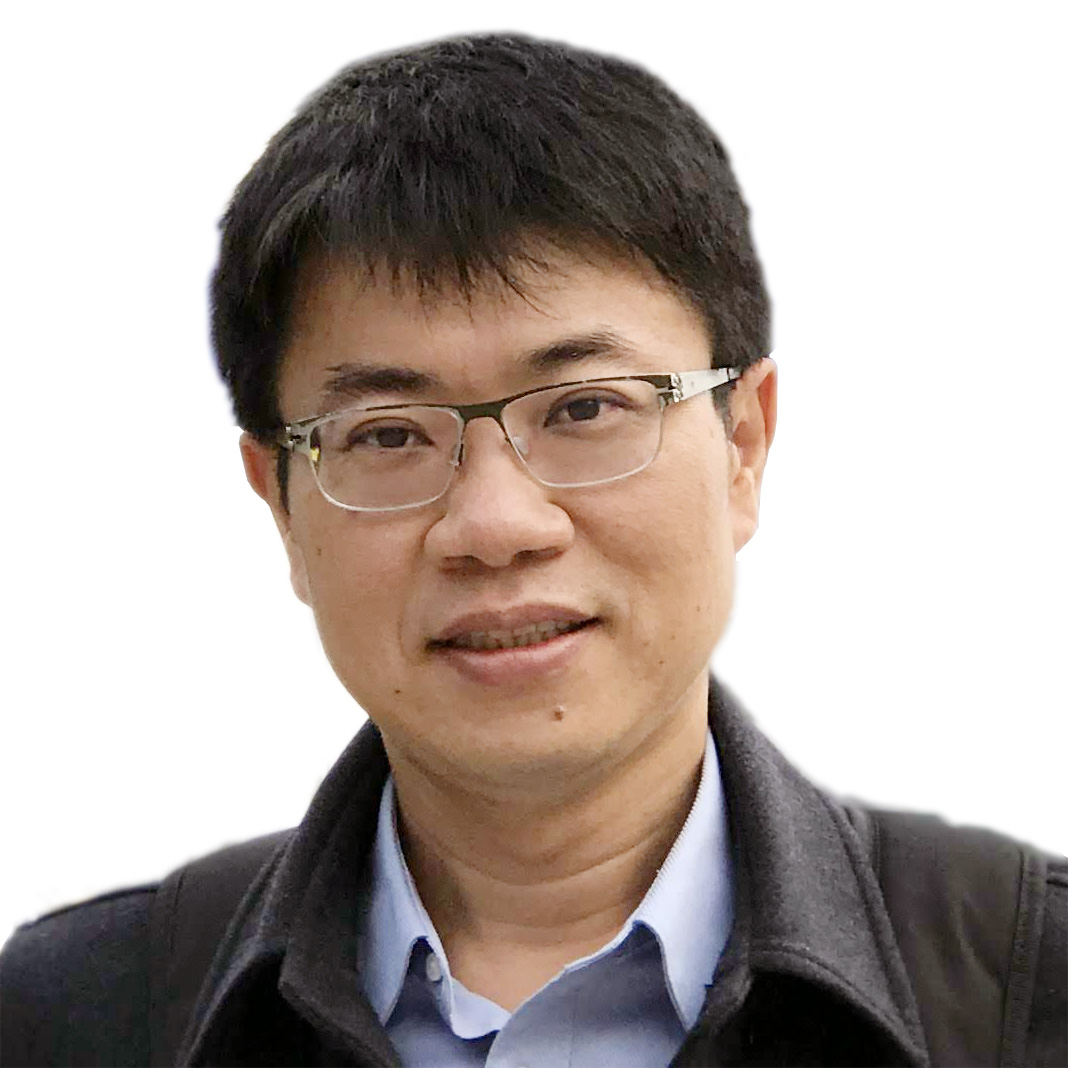}}]{Chih-Chyau Yang}
received the B.S. degree in electrical engineering from National Cheng-Kung University (NCKU), Taiwan in 1996, and the M.S. degree in electronics engineering from National Chiao-Tung University, Hsinchu (NCTU), Taiwan in 1999. He is currently a principal engineer at Taiwan Semiconductor Research Institute (TSRI), Taiwan. His research interests include VLSI design, computer architecture, and platform-based SoC design methodologies.
\end{IEEEbiography}
\begin{IEEEbiography}[{\includegraphics[width=1in,height=1.25in,clip,keepaspectratio]{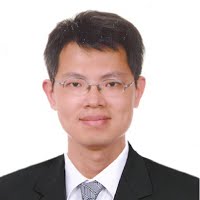}}]{Tian-Sheuan Chang}
	(S’93–M’06–SM’07)
	received the B.S., M.S., and Ph.D. degrees in electronic engineering from National Chiao-Tung University (NCTU), Hsinchu, Taiwan, in 1993, 1995, and 1999, respectively. 
	
	From 2000 to 2004, he was a Deputy Manager with Global Unichip Corporation, Hsinchu, Taiwan. In 2004, he joined the Department of Electronics Engineering, NCTU (as National Yang Ming Chiao Tung University (NYCU) in 2021), where he is currently a Professor. In 2009, he was a visiting scholar in IMEC, Belgium. His current research interests include system-on-a-chip design, VLSI signal processing, and computer architecture.
	
	Dr. Chang has received the Excellent Young Electrical Engineer from Chinese Institute of Electrical Engineering in 2007, and the Outstanding Young Scholar from Taiwan IC Design Society in 2010. He has been actively involved in many international conferences as an organizing committee or technical program committee member.
\end{IEEEbiography}

\end{document}